\documentclass[11pt,a4paper]{article}
\usepackage{jheppub}
\usepackage{subcaption}
\usepackage{setspace,braket}
\usepackage{amsmath, amssymb, bm}
\usepackage{etoolbox}

    \makeatletter
    \patchcmd{\maketitle}{\@fpheader}{}{}{}
\makeatother
\def\be{\begin{equation}}
\def\ee{\end{equation}}
\def\ba{\begin{eqnarray}}
\def\ea{\end{eqnarray}}

\def\({\left(}
\def\){\right)}
\def\[{\left[}
\def\]{\right]}

\newcommand{\bea}{\begin{eqnarray}}
\newcommand{\eea}{\end{eqnarray}}

\numberwithin{equation}{section}
\renewcommand{\thesection}{\arabic{section}}

\begin{document}
\renewcommand{\thefootnote}{\fnsymbol{footnote}}

\title{From singular to regular: revisiting thermodynamics of Bardeen-AdS black holes}
\author[a,b]{Meng-Sen Ma,\footnote{Corresponding author. E-mail address: mengsenma@gmail.com}}
\author[b]{Yun He,}
\author[a,b]{Xiao-Ming Wang,}
\author[b,c]{Huai-Fan Li}
\affiliation[a]{Department of Physics, Shanxi Datong
University,  Datong 037009, China}
\affiliation[b]{Institute of Theoretical Physics, Shanxi Datong
University, Datong 037009, China}
\affiliation[c]{College of General Education, Shanxi College of Technology, Shuozhou 036000, China}

\abstract{
We explore the thermodynamic properties of the regular Bardeen-AdS black hole obtained by imposing an additional constraint on a singular ``mother" black hole. This constraint eliminates the physical singularity but leads to the breakdown of the standard first law of black hole thermodynamics. The singular black hole exhibits a reentrant phase transition similar to that of the higher-dimensional Kerr-AdS black hole. The Bardeen-AdS black hole exhibits $P-V$ criticalities similar to that of the RN-AdS black hole, however has striking differences in its Gibbs free energy behavior. In particular, the characteristic “swallow-tail” structure associated with first-order phase transitions disappears. Instead, an ``8-shaped" or ``c-shaped" structure occurs, signifying a first-order phase transition or a zeroth-order phase transition between the small black hole and the large black hole phases, respectively. Our analysis suggests that constraint-induced modifications in the thermodynamic phase space may have deep consequences for the critical behaviors of black holes.

}
\maketitle
\onehalfspace

\renewcommand{\thefootnote}{\arabic{footnote}}
\setcounter{footnote}{0}
\section{Introduction}
\label{intro}

The presence of curvature singularities has long been regarded as one of the most fundamental shortcomings of classical general relativity. In these regions, spacetime curvature diverges, and the known laws of physics cease to apply. Resolving such singularities is generally expected to require a full quantum theory of gravity. However, interestingly, a number of nonsingular or “regular” black hole solutions have been proposed within classical or semiclassical frameworks. The first and the most influential example of regular black holes is the Bardeen solution, which was proposed by Bardeen in 1968\cite{Bardeen.87.1968}, and later was shown to arise from general relativity coupled to a nonlinear magnetic monopole\cite{AyonBeato.149.2000}. Regular black holes typically involve non-standard matter sources\cite{Dymnikova.235.1992, Hayward.031103.2006, Dymnikova.1545002.2015} or modifications to the energy-momentum tensor, such as nonlinear electrodynamics (NED)\cite{AyonBeato.5056.1998, Bronnikov.044005.2001, Burinskii.104017.2002, Dymnikova.4417.2004, Breton.643.2005, Berej.885.2006, Balart.124045.2014, Ma.529.2015, Rodrigues.024062.2016, Fan.124027.2016, Nojiri.104008.2017, Ghosh.104050.2018, Gulin.025015.2018, Bokulic.124059.2021,Maeda.108.2022, Li.104046.2024}, exotic scalar fields\cite{Bronnikov.251101.2006}, or noncommutative geometry-inspired models\cite{Nicolini.547.2006, Nicolini.1229.2009}. For more details, one can refer to the recent review article\cite{Lan.202.2023}.


It has long been known that for many regular black holes there is an inconsistency among the temperature, the entropy and the first law of thermodynamics\footnote{To our knowledge, only two types of regular black holes can avoid this issue. The first one is the $(2+1)$-dimensional regular black holes\cite{Ma:2017}, and the other one is the regular black hole constructed from pure gravity\cite{Bueno.139260.2025}. Because there are naturally no additional constraints among the thermodynamic quantities for these regular black holes.}\cite{WenJuan.453.2006, Ma.245014.2014}. In previous studies on the thermodynamics of regular (non-singular) black holes, it is often assumed that the first law of thermodynamics holds. Based on this assumption, researchers typically proceed in one of two ways: either they take the temperature to be the Hawking temperature and derive the entropy accordingly, or they assume that the black hole entropy obeys the Bekenstein–Hawking area law and deduce the temperature from it. However, these two approaches are generally incompatible for regular black holes—one must choose between them, inevitably abandoning the other\cite{Myung.012.2007, Banerjee.124035.2008, Nicolini.097.2011, Akbar.070401.2012, Smailagic.1350010.2013, Zhang.145007.2018, Li.1950336.2019, Tzikas.219.2019,Guo.025402.2022, Golchin.066007.2022,Simovic.044029.2024, Bakopoulos.L101502.2024, Cisterna.2025.23467}.

In this work, we re-explore in detail the thermodynamic properties of the Bardeen-AdS black hole on the basis of a new idea\cite{Ma.2507.09551}. We first construct a singular black hole, which can be regarded as the ``mother" black hole of the Bardeen-AdS black hole. The ``mother" singular black hole satisfies the standard first law of black hole thermodynamics. We analyze its thermodynamic behavior by calculating key quantities such as the temperature, entropy, specific heat, and Gibbs free energy. After imposing an additional constraint on the black hole parameters, we can obtain the Bardeen-AdS black hole from the singular black hole solution. However, the imposition of this constraint reduces the dimensionality of the thermodynamic phase space and introduces dependence among black hole parameters. As a result, the standard form of the first law of black hole thermodynamics no longer holds. While the temperature of the Bardeen-AdS black hole can still be computed via the surface gravity, other thermodynamic quantities such as specific heat and Gibbs free energy cannot be directly obtained from a consistent first law  for the Bardeen-AdS black hole.

To address this issue, we apply the constraint to project the thermodynamic quantities derived in the ``mother" singular black hole into the reduced thermodynamic phase space associated with the Bardeen-AdS black hole. This procedure ensures consistency between the temperature and the thermodynamic potentials and allows us to analyze the phase structure of the Bardeen-AdS black hole in a physically meaningful way.

Recently, a Hamiltonian formulation of black hole thermodynamics has been proposed in\cite{Baldiotti:2016lcf, Baldiotti201722}, where the thermodynamic variables are treated as canonical variables in an extended phase space. In that framework, new degrees of freedom are introduced to ensure the homogeneity of thermodynamic variables and to construct an effective temperature. Although this approach provides a consistent Hamiltonian description of the AdS black hole thermodynamics, the resulting temperature is an effective one rather than the genuine Hawking temperature determined by the surface gravity. Our present method follows a conceptually opposite route: instead of enlarging the phase space by introducing new thermodynamic variables, we start from a higher-dimensional thermodynamic phase space associated with a singular “mother” black hole, and then reduce it by imposing a specific constraint. This constraint removes the curvature singularity and yields the regular Bardeen-AdS black hole, but at the cost of breaking the standard first law. From this perspective, the breakdown of the first law is a direct manifestation of the constraint-induced reduction of phase space, rather than an inconsistency of the thermodynamic framework.

The paper is arranged as follows. In section 2 we first derive the `` mother" singular black hole by taking a nonlinear electrodynamics source as gravitational sources. In section 23 we calculate the thermodynamic quantities of the ``mother" black hole and analyze its critical behaviors and phase structure. We then impose the additional constraint to obtain the Bardeen-AdS black hole in section 4, and study its thermodynamic properties. At last, we summarize our results and discuss the possible future study in section 5.

\section{The singular ``mother"  black hole}
In this part, we will construct a singular black hole in the framework of the general relativity with the nonlinear electrodynamics source. The Einstein field equation with the cosmological constant has the form
\be
G_{\mu\nu}+\Lambda g_{\mu\nu}=8\pi T_{\mu\nu}.
\ee
We consider the static spherically symmetric solution, which has the following metric ansatz
\be\label{staticmetric}
ds^2=-f(r)dt^2+f(r)^{-1}dr^2 + r^2d\Omega^2.
\ee
For simplicity, we take the metric function in the following form,
\be\label{efm}
f(r)=1-\frac{2m(r)}{r}.
\ee
Substituting it into the field equation, one can obtain
\be\label{mr}
-\frac{ 2m'(r)}{r^2}+\Lambda=8\pi T^0_{~0}.
\ee
After integration, we have
\be\label{mr2}
m(r)=M+4\pi\int_{r}^{\infty}r^2T^0_{~0}dr+\frac{\Lambda}{6}r^3,
\ee
where $M$ is an integration constant, which is just the ADM mass in the asymptotically flat spacetime.

We consider a nonlinear electrodynamic matter fields as the gravitational source, whose Lagrangian takes the form
\be\label{Bardeen_L}
\mathcal{L(F)}=-\alpha  \left(\frac{\sqrt{F/2}}{1+\sqrt{F/2}}\right)^{5/2}.
\ee
It includes only one parameter $\alpha$, which is a positive coupling constant with the dimension of $[L]^{-2}$. $F$ is a scalar field and is defined as $F=F_{\mu\nu}F^{\mu\nu}$. 

In the spherically symmetric case with pure magnetic fields, $F_{\mu\nu}$ involves a radial magnetic field $F_{23}$ and satisfies
\be
F_{23}=Q_m \sin\theta,
\ee
where $Q_m$ is the magnetic charge. Thus $F=2F_{23}F^{23}=\frac{2Q_m^2}{r^4}$. 

The energy-momentum tensor derived from $L(F)$ is
\be
T_{\alpha\beta}=g_{\alpha\beta}\mathcal{L}+4\mathcal{L}_FF_{\alpha\mu}F^{\mu}_{~\beta},
\ee
where $\mathcal{L}_F\equiv \partial{\mathcal{L}}/\partial{F}$. In the static spherically symmetric case, we have $T^0_{~0}=\mathcal{L}$.

Now, from Eq.(\ref{mr2}) we can obtain
\be
m(r)=M-\frac{4\pi}{3} \alpha  Q_m^{3/2}\left[1-\frac{r^3}{\left(Q_m+r^2\right)^{3/2}}\right]+\frac{\Lambda}{6}r^3.
\ee
Therefore, the metric function takes the form of
\be\label{SBH_metric}
f(r)=1-\frac{2(M-\frac{4\pi}{3} \alpha  Q_m^{3/2})}{r}-\frac{8\pi}{3} \alpha  Q_m^{3/2}\frac{r^2}{\left(Q_m+r^2\right)^{3/2}}-\frac{\Lambda}{3}r^2.
\ee

\section{Thermodynamics of the singular ``mother" black hole}

We now try to study the thermodynamic properties of this singular black hole.
The Hawking temperature can be calculated directly from the metric function,
\be\label{T_SBH}
T=\frac{f'(r_{+})}{4\pi}=\frac{1-\Lambda r_{+}^2}{4\pi r_{+}}+2\alpha r_{+}\left(\frac{Q_m}{Q_m+r_{+}^2}\right)^{5/2}.
\ee
Irrespective of the values of $(Q_m, \Lambda, \alpha)$, the temperature always diverges in the limit $r_{+}\rightarrow 0$.

In this work, we will consider the extended phase space, where the dimensional parameters, such as the cosmological constant and the parameter $\alpha$, are both included in the thermodynamic phase space of the black hole\cite{Kastor.195011.2009}. Especially, the cosmological constant is treated as the pressure $P=-\Lambda/8\pi$\cite{Dolan.235017.2011}. One can easily check that the first law of black hole thermodynamics is satisfied,
\be
dM=TdS+\Phi dQ_m+\mathcal{A}d\alpha +VdP,
\ee
where $S=A/4$ and $\Phi, \mathcal{A}, V$ are the conjugated quantities of $Q_m, \alpha, P$, respectively. Besides, they also fulfill the Smarr relation:
\be
M/2=TS+\Phi Q_m-\mathcal{A}\alpha -VP.
\ee

From Eq.(\ref{T_SBH}) we can get the pressure as a function of $(T,r_{+}, \alpha, Q_m)$,
\be
P=\frac{T}{2 r_{+}}-\frac{1}{8 \pi  r_{+}^2}+\alpha  \left(\frac{Q_m}{Q_m+r_{+}^2}\right)^{5/2}.
\ee
Compared with the Van der Waals equation, the specific volume should be $v=2r_{+}$. For simplicity, below we consider the $P-r_{+}$ criticality instead.

\begin{figure}[!htb]
	\centering{
	\includegraphics[width=7.5cm]{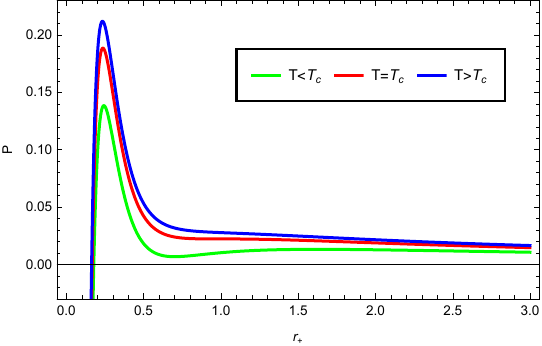} \hspace{0.5cm}
        \includegraphics[width=7.5cm]{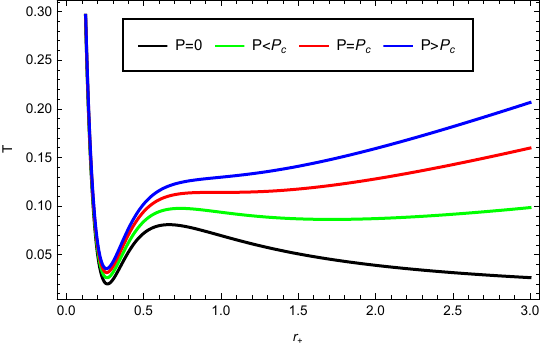} 
         \caption{The $P-r_{+}$ and the $T-r_{+}$ curves of the singular black hole for fixed $Q_m=0.1$ and $\alpha=2$.  } \label{fig_PT_SBH}
	}
\end{figure}

Now we can study the critical behaviors of the black hole. If critical points exist, they can be derived from
\be
\frac{\partial P}{\partial r_{+}}=0, \quad \frac{\partial^2 P}{\partial r_{+}^2}=0.
\ee
The algebraic equations are too complicated to obtain analytic solutions. We set $Q=0.1$ and $\alpha=2$ to derive a set of numerical solutions:
\be
T_c=0.114, \quad r_{+c}=0.979, \quad P_c=0.0222.
\ee
Its $P-r_{+}$ criticality is shown in Fig.\ref{fig_PT_SBH}. Compared with the RN-AdS black hole, this black hole exhibits an additional branch in the small black hole region. Its behavior is similar to that of the higher-dimensional rotating-AdS black holes\cite{Altamirano.101502.2013} and the Gauss-Bonnet-AdS black hole\cite{Wei.044057.2014}. In Fig.\ref{fig_PT_SBH}, we also depict the $T-r_{+}$ curves. Only if $P>0$, the temperature will tend to infinity as $r_{+} \rightarrow \infty$. When $P<P_c$, the temperature also exhibits four branches, corresponding to the $P-r_{+}$ curve in the $T<T_c$ case. 


Next, we examine the heat capacity, which not only characterizes the local thermodynamic stability but also provides insights into the microscopic degrees of freedom of the black hole. We calculate the heat capacity at constant $(Q_m, P, \alpha)$,
\be\label{C_SBH}
C=\left.\frac{\partial M}{\partial T}\right|_{Q_m,P,\alpha}.
\ee

\begin{figure}[!hbt]
	\centering{
	\includegraphics[width=8cm]{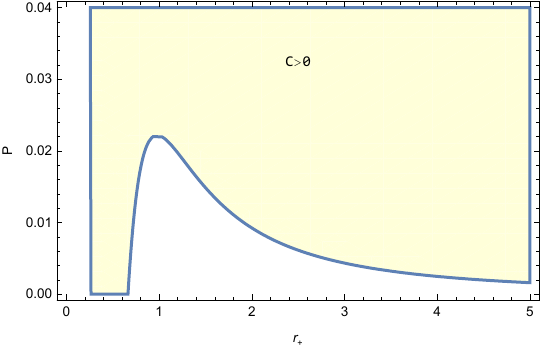} 
         \caption{The region with $C>0$ for fixed $Q_m=0.1$ and $\alpha=2$. The white areas correspond to $C<0$. } \label{fig_C0_SBH}
	}
\end{figure}

\begin{figure}[!hbt]
	\centering
    \begin{subfigure}{.45\textwidth}
    \centering
    \includegraphics[width=7cm]{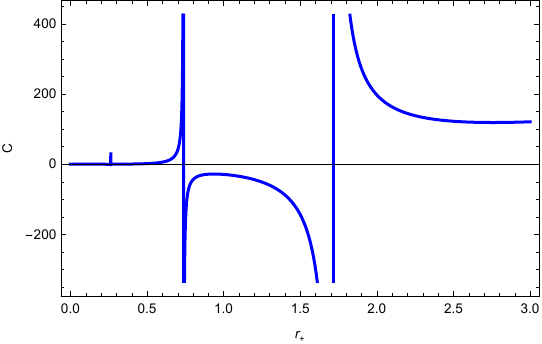}
    \caption{$P=0.012<P_c$}
    \label{fig_C1_SBH}
    \end{subfigure}
    \begin{subfigure}{.45\textwidth}
    \centering
    \includegraphics[width=7cm]{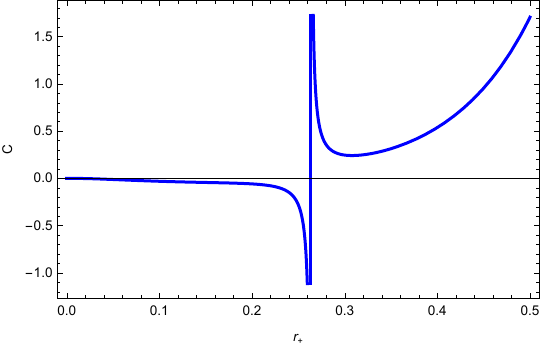}
    \caption{Magnification of (a) in the region $[0,0.5]$}
    \label{fig_C1b_SBH}
    \end{subfigure}
    \begin{subfigure}{.45\textwidth}
    \centering
         \includegraphics[width=7cm]{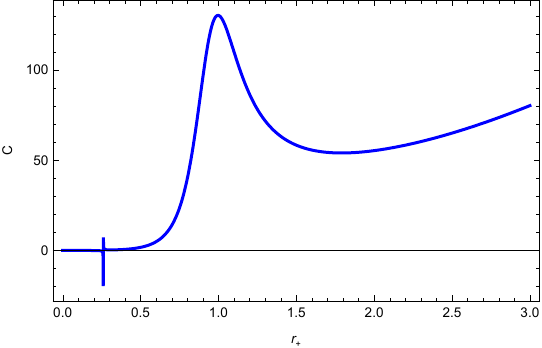}
    \caption{$P_c<P=0.025<P_0$}
    \label{fig_C3_SBH}
    \end{subfigure}
     \begin{subfigure}{.45\textwidth}
    \centering
         \includegraphics[width=7cm]{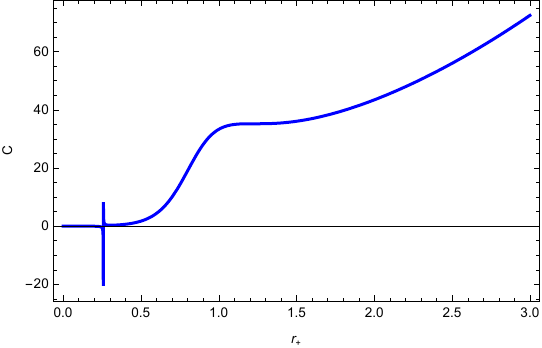}
    \caption{$P=P_0$}
    \label{fig_C2_SBH}
    \end{subfigure}
     \caption{The behaviors of $C-r_{+}$ for different values of $P$ for fixed $Q_m=0.1$ and $\alpha=2$. Here $P_0=0.0354$. } \label{fig_C_SBH}
\end{figure}

For fixed $Q_m$ and $\alpha$, there will be one or two regions in which the heat capacity is positive, depending on the values of $P$. This is shown in Fig.\ref{fig_C0_SBH}. 

The behaviors of the heat capacity are depicted in Fig.\ref{fig_C_SBH}. When $P<P_c$, the heat capacity has three divergent points, which correspond to the three extrema of the temperature. From left to right, in the second and fourth regions, the heat capacity is positive, which is a sign of local thermodynamic stability.  

\begin{figure}
	\centering{
     \begin{subfigure}{.45\textwidth}
    \centering
    \includegraphics[width=7cm]{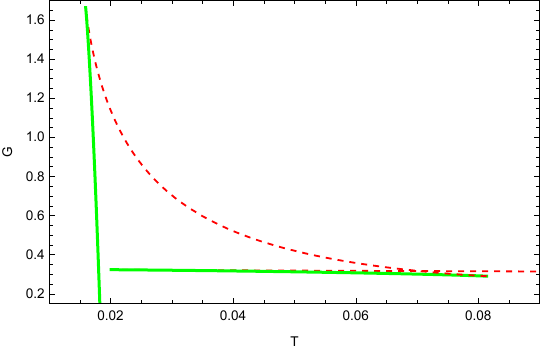}
    \caption{$P=0.0004<P_t$}
    \label{fig_G1_SBH}
    \end{subfigure}
    \begin{subfigure}{.45\textwidth}
    \centering
    \includegraphics[width=7cm]{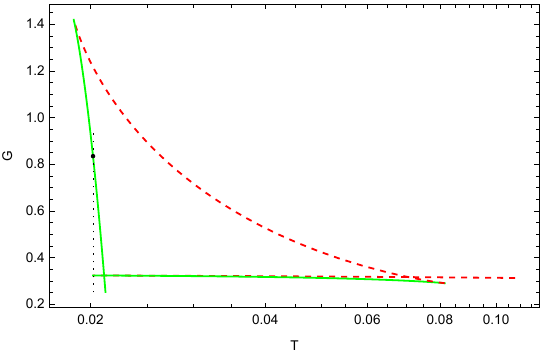}
    \caption{$P_t<P=0.00055<P_z$}
    \label{fig_G2_SBH}
    \end{subfigure}\vspace{0.5cm}
    \begin{subfigure}{.45\textwidth}
    \centering
    \includegraphics[width=7cm]{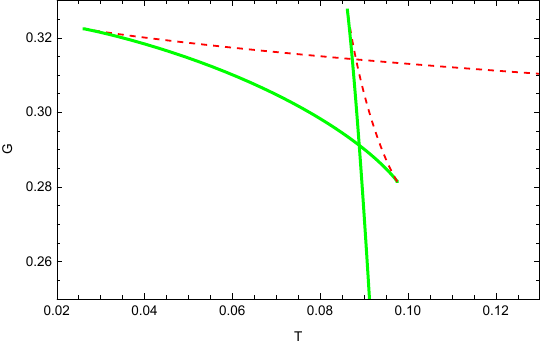}
    \caption{$P_z<P=0.012<P_c$}
    \label{fig_G3_SBH}
    \end{subfigure}
    \begin{subfigure}{.45\textwidth}
    \centering
    \includegraphics[width=7cm]{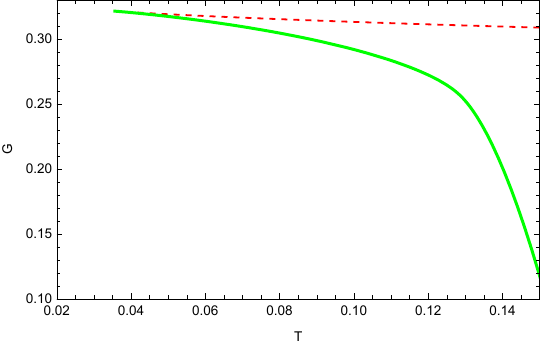}
    \caption{$P>P_c$}
    \label{fig_G4_SBH}
    \end{subfigure}
    \caption{The $G-T$ curves for different values of $P$ for fixed $Q_m=0.1$ and $\alpha=2$. Here $P_t=0.00047,~P_z=0.00064$. The red dashed lines correspond to the phases with negative heat capacity.  } \label{fig_G_SBH}
	}
\end{figure}

When $P>P_c$, the two divergent points on the right side disappear(Fig.\ref{fig_C3_SBH} and Fig.\ref{fig_C2_SBH}). There exists a particular pressure, denoted as $P_0$. When $P<P_0$, the heat capacity develops a peak. This behavior resembles the Schottky anomaly and may indicate the presence of discrete energy levels in the underlying microstructure of the black hole\cite{Dinsmore.054001.2020}. For pressures $P\geq P_0$, the peak vanishes. Interestingly, when $P=P_0$, the heat capacity exhibits two platforms (Fig.\ref{fig_C3_SBH}), reminiscent of the heat capacity behavior in a diatomic ideal gas. This analogy further supports the interpretation that the black hole may possess discrete energy levels and internal microscopic degrees of freedom.

According to the first law, the parameter $M$ now plays the role of the enthalpy\cite{Kastor.195011.2009}. In the fixed $(Q_m, P, \alpha)$ ensemble, we can define the Gibbs free energy as
\be\label{G_SBH}
G=M-TS=\frac{1}{6} \left[4 \pi  r_+^3 \left[\frac{\alpha}{Q}\left(Q-2 r_+^2\right) \left(\frac{Q}{Q+r_+^2}\right){}^{5/2}-P\right]+8 \pi  \alpha  Q^{3/2}+\frac{3 r_+}{2}\right].
\ee
which is depicted in Fig.\ref{fig_G_SBH}. There are another two special pressures, $P_t$ and $P_z$. When $P<P_t$, only the larger black hole phase exists. When $P_t<P<P_{z}$, a zeroth-order phase transition may occur. Initially, at low temperatures, the black hole is thermodynamically favored in the larger black hole phase. As the temperature increases, a zeroth-order phase transition drives it into the smaller black hole phase, followed by a first-order phase transition that restores the larger black hole phase at higher temperatures. This is characteristic of the reentrant phase transition\cite{Altamirano.101502.2013,  Frassino.080.2014, Hennigar.80568072.2015, Zou.256.2017, Ma.1257631.2018}. When $P_z<P<P_c$, the ``swallow tail" behavior manifests that it is a first-order phase transition between the smaller black hole and the larger black hole phases.

\section{Thermodynamics of the Bardeen-AdS black hole}
First, we should derive the regular Bardeen-AdS black hole from its ``mother" singular black hole.  This can be easily realized by adding the extra constraint,
\be\label{Bardeen_cons1}
M=\frac{4\pi}{3} \alpha  Q^{3/2},
\ee
or equivalently
\be\label{Bardeen_cons2}
\alpha=\frac{3 }{8 \pi  r_{+}^2}\left(\frac{Q+r_{+}^2}{Q}\right)^{3/2},
\ee
 on the metric function, Eq.(\ref{SBH_metric}).
 
In this way, we obtain the Bardeen-AdS black hole,
\be\label{Bardeen_metric}
f(r)=1-\frac{8\pi}{3} \frac{\alpha  Q^{3/2}r^2}{\left(Q+r^2\right)^{3/2}}-\frac{\Lambda}{3}r^2=1-\frac{2 M r^2}{\left(Q+r^2\right)^{3/2}}-\frac{\Lambda}{3}r^2.
\ee
By calculating the Kretschmann scalars, one can verify that it is indeed a regular black hole solution. When $\Lambda=0$, it is just the Bardeen regular black hole.

\begin{figure}[!hbt]
	\centering{
	\includegraphics[width=7cm]{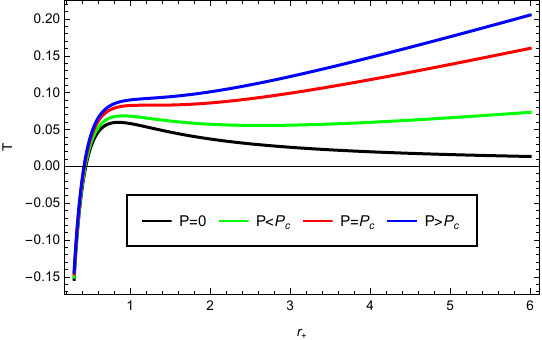} \hspace{0.5cm}
        \includegraphics[width=7cm]{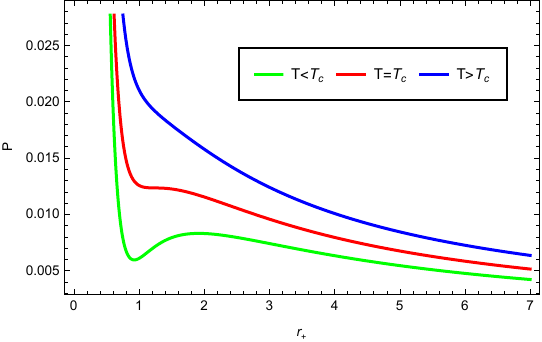}
         \caption{The $T-r_{+}$ and $P-r_{+}$ curves of the Bardeen-AdS black hole with fixed $Q_m=0.1$.  } \label{fig_TP_RBH}
	}
\end{figure}

The temperature of this regular black hole can be derived from the metric function (\ref{Bardeen_metric}), or can be obtained directly by adding the constraint (\ref{Bardeen_cons2}) on Eq.(\ref{T_SBH}),
\be
\text{Eq.}(\ref{T_SBH})\xrightarrow{(\ref{Bardeen_cons2})}T=\frac{1}{4 \pi  r_{+}}\left(1-\Lambda  r_{+}^2-\frac{3 Q_m}{Q_m+r_{+}^2}\right).
\ee
When $r_{+} \rightarrow 0$, for any finite $(Q, \Lambda)$ the temperature tends to $T \sim -\frac{1}{2\pi r_{+}} \rightarrow -\infty$. Therefore, we expect that the temperature of the Bardeen-AdS black hole has different behaviors from that of its  singular ``mother" black hole. From this temperature, we can derive the $P$ as a function of $(T, r_{+}, Q_m)$,
\be
P=\frac{T}{2 r_+}+\frac{2Q_m-r_{+}^2}{8 \pi r_+^2 \left( Q_m+  r_+^2\right)}.
\ee

In Fig.\ref{fig_TP_RBH}, we depict the curves of $T-r_{+}$ and $P-r_{+}$. Outwardly, it seems that the Bardeen-AdS has similar critical behaviors to that of the RN-AdS black hole\cite{Kubiznak.033.2012}. We can also calculate the critical point for the fixed $Q_m=0.1$,
\be
T_c=0.0828, \quad r_{+c}=1.241, \quad P_c=0.0122.
\ee

However, as we discussed in \cite{Ma.2507.09551}, the introduction of the extra constraint renders the thermodynamic variables no longer independent, thereby leading to a breakdown of the first law of black hole thermodynamics. One cannot define the heat capacity according to the first law. To obtain the heat capacity of the Bardeen-AdS black hole, the only feasible approach is to put the constraint (\ref{Bardeen_cons2}) on the corresponding heat capacity of the singular black hole. The result is
\be\label{C_RBH}
\text{Eq.}(\ref{C_SBH})\xrightarrow{(\ref{Bardeen_cons2})}C=\frac{8 \pi ^2 r_{+}^3 \left(Q+r_{+}^2\right)^2 T}{Q^2 \left(8 \pi  P r_{+}^2-4\right)+2 Q r_{+}^2 \left(8 \pi  P r_{+}^2+5\right)+r^4 \left(8 \pi  P r_{+}^2-1\right)}.
\ee

\begin{figure}[!hbt]
	\centering{
    \begin{subfigure}{.45\textwidth}
    \centering
    \includegraphics[width=7cm]{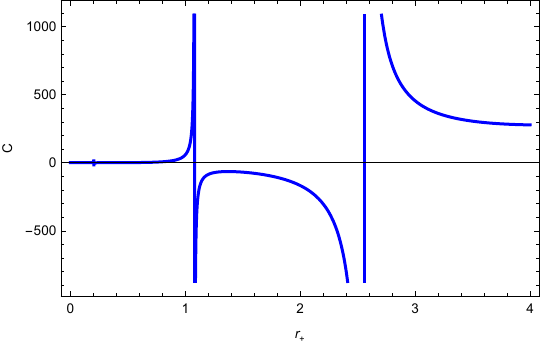}
    \caption{$P=0.005<P_1=0.0092$}
    \label{fig_C1_RBH}
    \end{subfigure}
    \begin{subfigure}{.45\textwidth}
    \centering
    \includegraphics[width=7cm]{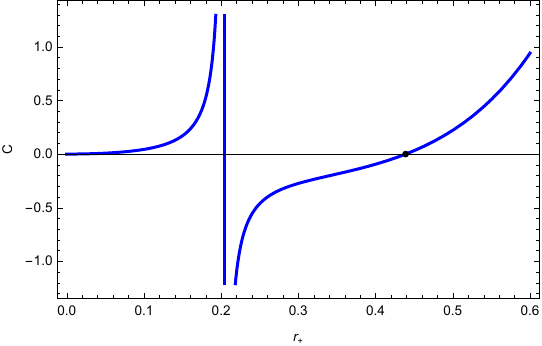}
    \caption{Magnification of (a) in the region $[0,0.6]$}
    \label{fig_C2_RBH}
    \end{subfigure} \vspace{0.5 cm}
    
    \begin{subfigure}{.45\textwidth}
    \centering
         \includegraphics[width=7cm]{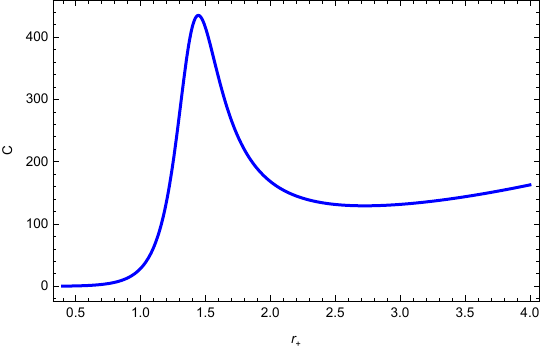}
    \caption{$P_1<P=0.01<P_0$}
    \label{fig_C3_RBH}
    \end{subfigure}
     \begin{subfigure}{.45\textwidth}
    \centering
         \includegraphics[width=7cm]{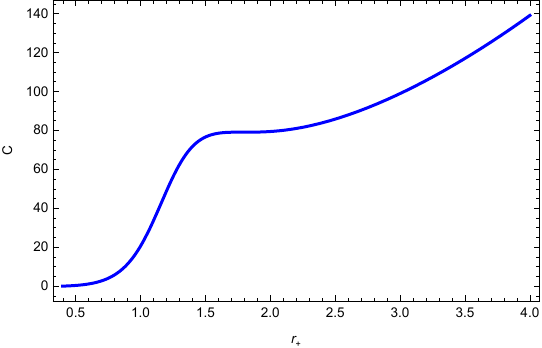}
    \caption{$P=P_0=0.0147$}
    \label{fig_C4_RBH}
    \end{subfigure}
     \caption{The behaviors of $C-r_{+}$ for the Bardeen-AdS black holes for fixed $Q_m=0.1$. In (b), the black dot corresponds to the zero temperature. On the left of this point, the temperature is negative. } }\label{fig_C_RBH}
\end{figure}

Because, in this case, 
\be
C \neq \frac{\partial M}{\partial T} =\frac{\partial M/\partial r_{+}}{\partial T/\partial r_{+}},
\ee
the extrema of the temperature do not correspond to the divergent point of the heat capacity. To derive the divergent points, we set $r_{+}^2=x$ in the denominator of Eq.(\ref{C_RBH}), 
\be
x \left(8 \pi  P Q^2+10 Q\right)+x^2 (16 \pi  P Q-1)+8 \pi  P x^3-4 Q^2=0.
\ee
According to the discriminant of cubic algebraic equations, we obtain
\be
\Delta=2560 \pi ^3 P^3 Q^3+15312 \pi ^2 P^2 Q^2+2376 \pi  P Q-7.
\ee
For $Q=0.1$, we find that it will have three real roots when $P<P_1=0.0092$. This is depicted in Fig.\ref{fig_C1_RBH} and Fig.\ref{fig_C2_RBH}. The first divergent point lies in the region where the temperature is less than zero, and thus should be excluded.  When $P>P_1$, in the area with a positive temperature, the heat capacity is always positive and no divergent point exists(Fig.\ref{fig_C3_RBH} and Fig.\ref{fig_C4_RBH}). When $P<P_0$, the heat capacity also exhibits a Schottky anomaly-like behavior.

The Gibbs free energy is usually defined according to the first law of thermodynamics and the Legendre transformation. The breakdown of the first law means that we cannot define the Gibbs free energy of the Bardeen-AdS black hole as usual directly. We take the same approach by adding the constraint on the Gibbs free energy of the singular black hole, and obtain
\be
\text{Eq.}(\ref{G_SBH})\xrightarrow{(\ref{Bardeen_cons2})}G=\frac{1}{6} \left[\frac{3 \left(Q+r_+^2\right){}^{3/2}}{r_+^2}+\frac{3 r_+ \left(Q-2 r_+^2\right)}{2 \left(Q+r_+^2\right)}+\frac{3 r_+}{2}-4 \pi  P r_+^3\right].
\ee

\begin{figure}[!hbt]
	\centering{
	\includegraphics[width=9cm]{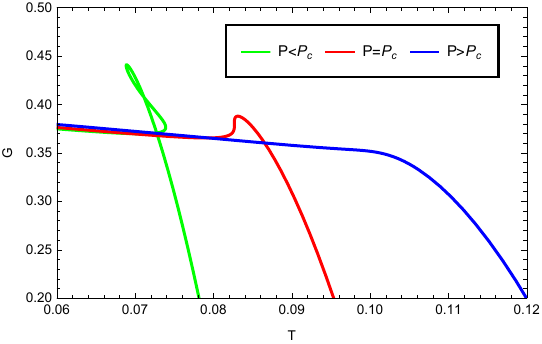} 
        \caption{The $G-T$ curves for fixed $Q_m=0.1$ by taking $P_c$ as the benchmark.  } \label{fig_G1_RBH}
	}
\end{figure}

As is shown in Fig.\ref{fig_G1_RBH}, when $P\geq P_c$ the Gibbs free energy has only one branch, and when $P<P_c$ it also has three branches corresponding to the temperature. Obviously, the behaviors of $G-T$ curves are different from those of the RN-AdS black hole. Especially, when $P<P_c$, there is not a `` swallow tail ", but a ``8-shaped" knot.

\begin{figure}[!hbt]
	\centering{
    \begin{subfigure}{.45\textwidth}
    \centering
    \includegraphics[width=7cm]{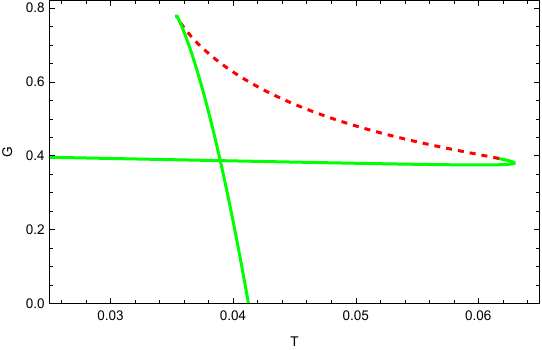}
    \caption{$P=0.002$}
    \label{fig_G2_RBH}
    \end{subfigure}
    \begin{subfigure}{.45\textwidth}
    \centering
    \includegraphics[width=7cm]{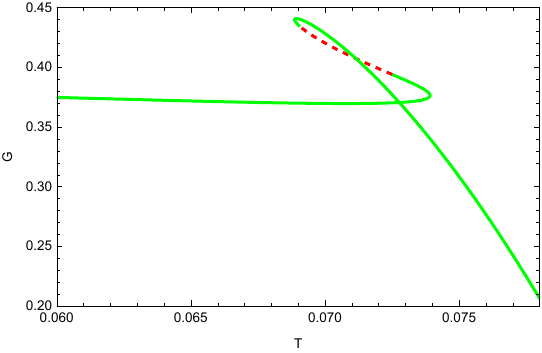}
    \caption{$P=0.008$}
    \label{fig_G3_RBH}
    \end{subfigure} \vspace{0.5 cm}
    
    \begin{subfigure}{.45\textwidth}
    \centering
         \includegraphics[width=7cm]{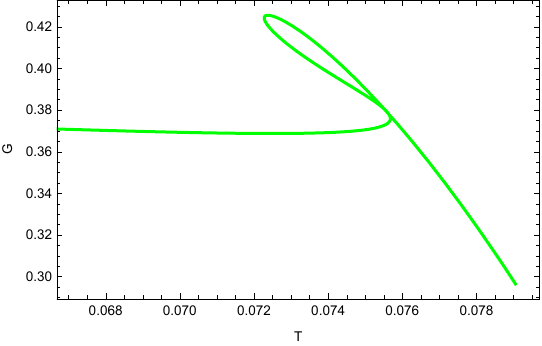}
    \caption{$P=P_z=0.0089$}
    \label{fig_G4_RBH}
    \end{subfigure}
     \begin{subfigure}{.45\textwidth}
    \centering
         \includegraphics[width=7cm]{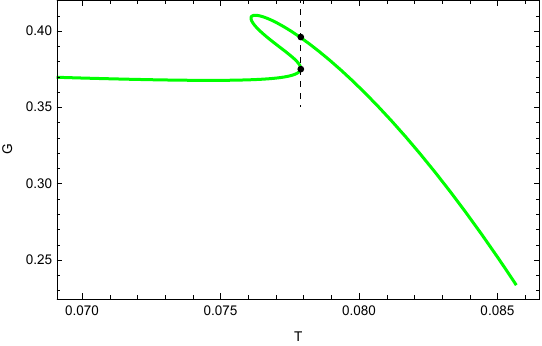}
    \caption{$P_z<P=0.01<P_c$}
    \label{fig_G5_RBH}
    \end{subfigure}
     \caption{The refined phase structures of the Bardeen-AdS black holes for fixed $Q_m=0.1$ when $P<P_c$. The red dashed curves represent the regions of negative heat capacities. } \label{fig_G_RBH}}
\end{figure}

In fact, the Bardeen-AdS black hole has more fruitful phase structures. When $P=0$, the larger black hole branch disappear, which has been discussed in\cite{Ma.2507.09551}. As is shown in Fig.\ref{fig_G_RBH}, there exists another special value of $P=P_z$, below which, but not the critical pressure $P_c$, the $G-T$ curves exhibit the `` 8 "-like shape. When $0<P\ll P_z$, the upper loop of the `` 8 " shape shrinks and the lower loop enlarges, which makes the `` 8 " shape look like a swallow tail(Fig.\ref{fig_G2_RBH}). It should be noted that, not the whole intermediate branch, but only part of which is locally thermodynamically unstable. This property is also different from that of the RN-AdS and other AdS black holes. At $P=P_z$, the lower loop of the `` 8 "-like shape disappear, and it becomes a `` 0 "-like shape. When $P_z<P<P_c$, the loop splits, evolving into a ``C-shaped" structure. In this case, a zeroth-order phase transition occurs between the smaller black hole and the larger one and no first-order phase transitions exist.

\section{Discussions and Conclusions}
\label{Conclusions}

In this work, we have investigated the thermodynamics of the regular Bardeen-AdS black hole and its singular “mother” black hole. The regular solution can be obtained from the singular one by imposing a specific constraint. This constraint eliminates the physical singularity, leading to a smooth black hole geometry. However, it also induces significant modifications to the thermodynamic structure. Because the constraint can leads to a loss of independence among the thermodynamic variables, causing the standard form of the first law of black hole thermodynamics to break down.

The singular black hole has rich phase structures, including both first-order and zeroth-order phase transitions. In particular, it displays a reentrant phase transition, reminiscent of the phenomena observed in higher-dimensional Kerr-AdS black holes. 

More interesting is the Bardeen-AdS black hole. we find that while it exhibits standard $P-V$ criticality akin to the RN-AdS black hole, its $G-T$ behavior departs significantly. In particular, the familiar swallow-tail structure, typically associated with first-order phase transitions, is absent. Instead, for pressures below a certain threshold $P_z$, the system undergoes a first-order phase transition with a Gibbs free energy diagram exhibiting a “8-shaped” structure. At $P=P_z$, this degenerates into a single-loop (“zero-like” shape). As the pressure increases further but remains below the critical pressure 
$P_c$, the diagram transforms into a “C-shaped” structure, characteristic of a zeroth-order phase transition between the small and large black hole phases. For 
$P>P_c$, the system enters a single-phase regime with no phase transitions.

This study highlights the importance of recognizing the thermodynamic consequences of constraints introduced on black holes. By identifying and calculating within the full thermodynamic phase space of the singular ``mother" black hole, we offer a consistent thermodynamic framework for understanding both the singular and regular black holes. Our results also suggest that regular black holes, though free from spacetime singularities, may exhibit thermodynamic behavior that is markedly distinct from their singular counterparts.

It would also be interesting to explore whether the Hamiltonian formalism\cite{Baldiotti201722} could be adapted to describe the constrained phase space of regular black holes. Such a study may provide a unified thermodynamic framework connecting the reduction and extension of black hole phase space.

\bigskip
\bigskip



\noindent\textbf{Declaration of competing interest}

The authors declare that they have no known competing financial interests or personal relationships that could have appeared to influence the work reported in this paper.

\bigskip
\noindent\textbf{Data availability}

No data was used for the research described in the article.

\acknowledgments
This work is supported in part by Shanxi Provincial Natural Science Foundation of China (Grant No. 202203021221211) and the Scientific and Technological Innovation Programs of Higher Education Institutions in Shanxi (Grant No. 2021L386). 

\bibliographystyle{JHEP}

\end{document}